\documentclass[showkeys,12pt,twocolumn]{article}

\usepackage{arxiv}
\usepackage[utf8]{inputenc}
\usepackage[backend=bibtex,style=phys,biblabel=brackets,articletitle=false]{biblatex}
\usepackage[english]{babel}
\usepackage[colorinlistoftodos, color=green!40, prependcaption]{todonotes}
\usepackage[colorlinks=true,linkcolor=blue,urlcolor=blue,citecolor=blue]{hyperref}
\usepackage{mathrsfs}
\usepackage{amsmath}
\usepackage{amsfonts}
\usepackage{amssymb}
\usepackage{amsthm}
\usepackage{mathdesign}
\usepackage{slashed}
\usepackage{mathtools}
\usepackage{xcolor}
\usepackage{enumitem}
\usepackage[capitalise]{cleveref}
\usepackage[normalem]{ulem}
\usepackage{float}
\usepackage{empheq}

\makeatletter
\newcommand{\mybf}[1]{%
    \begingroup
    \def\@tempa{#1}%
    \ifx\@tempa\@empty
        \textbf{#1}%
    \else
        \edef\@tempa{\noexpand\in@{#1}{abcdefghijklmnopqrstuvwxyzABCDEFGHIJKLMNOPQRSTUVWXYZ}}%
        \@tempa
        \ifin@
            \mathbf{#1}%
        \else
            \pmb{#1}%
        \fi
    \fi
    \endgroup
}
\makeatother

\newcommand\corr[1]{\textcolor{black}{#1}}

\begin{document}

\title{A note on thermal effects in non-linear models for plasma-based acceleration}

\author{
    Daniele Simeoni$^{1,*}$
\And Gianmarco Parise$^{2}$
\And Andrea Renato Rossi$^{3}$
\And Andrea Frazzitta$^{3,4}$
\And Fabio Guglietta$^{1}$
\And Mauro Sbragaglia$^{1}$ %
\and \\
$^{1}$Department of Physics \& INFN, Tor Vergata University of Rome, Via della Ricerca Scientifica 1, 00133, Rome, Italy
\and
$^{2}$INFN, Laboratori Nazionali di Frascati, Via Enrico Fermi 54, 00044, Frascati, Italy
\and
$^{3}$INFN, Section of Milan, via Celoria 16, 20133, Milan, Italy
\and
$^{4}$Department of Physics, La Sapienza University of Rome, Piazzale Aldo Moro 5, 00185, Rome, Italy
\and
$^{*}$\texttt{daniele.simeoni@roma2.infn.it}
}

\date{\today} 

\twocolumn[
\maketitle

\begin{abstract}
We investigate the impact of a non-negligible background temperature on relativistic plasma wakefields generated when a beam of charged particles passes through a neutral plasma at rest. \corr{We focus on the blowout regime, wherein the plasma response is highly non-linear: plasma electrons are radially blown out and expelled away from the propagation axis of the beam particles, creating a region (bubble) of ions without electrons. Our study builds upon earlier investigations for non-linear models of plasma wakefields developed in the limit of zero background temperature (Lu {\it et al.}, {\it Phys. Rev. Lett.} {\bf 96}, 165002 (2006)). In the presence of a non-zero background temperature, we characterize the model and focus on its predictions for the bubble size and the electromagnetic fields inside the bubble. Model predictions are studied in combination with PIC simulations.}
\end{abstract}
]

\section{Introduction \label{sec:intro}} 
\begin{figure}
    \centering
    \includegraphics[width=\columnwidth]{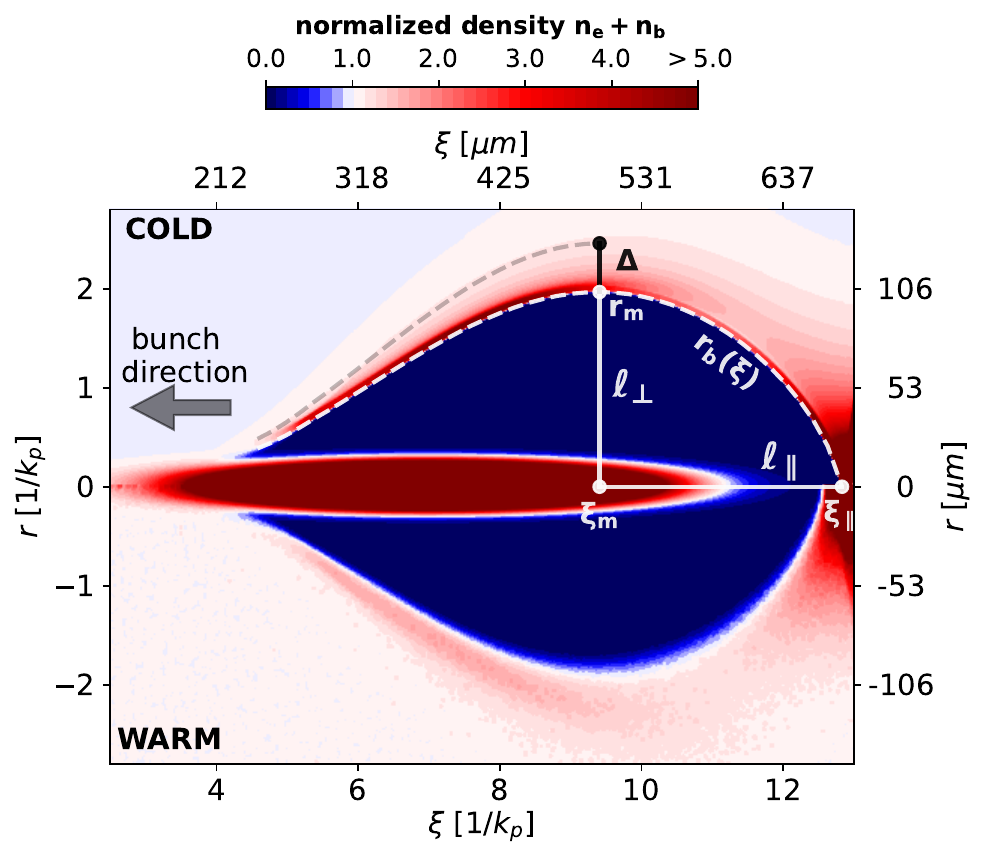}
    \caption{\corr{PIC snapshot of the first electron depletion bubble generated by an ultra-relativistic electron bunch propagating from right to left. The colormap represents the total normalized density ($n_b$ driver density, $n_e$ background plasma density), for two initial background temperatures:
    $k_B T_i = 0.00 \; \rm{keV}$ (top half of the figure) and $k_B T_i = 1.00 \; \rm{keV}$ (bottom half of the figure). The dashed white curve denotes the bubble boundary $r_b(\xi)$, the white solid lines denote its transverse and longitudinal extents, respectively $\ell_\bot$ and $\ell_\parallel$. The black line denotes the electron sheath $\Delta$.}
    }
    \label{fig:1}
\end{figure}

In plasma acceleration, a relativistic particle bunch or a short, intense laser pulse perturbs a plasma at rest, generating electromagnetic fields of order~$\lesssim \mathcal{O}(100)\; \rm{GV/m}$~\cite{blumenfeld-2007,gonsalves-2019} that can be used to accelerate and focus a witness of trailing beam particles properly injected in the plasma wake driven by the bunch/laser~\cite{tajima-1979,esarey-2009}. These two scenarios are commonly referred to as plasma wakefield acceleration (PWFA) and laser wakefield acceleration (LWFA). Theories have been developed to understand wakefield properties in the linear regime~\cite{katsouleas-1985,rosenzweig-1990,lu-2005}; however, when the plasma response becomes highly non-linear, a blowout regime is reached, wherein plasma electrons are radially blown out and expelled away from the propagating axis of the bunch/laser, due to the interaction with the charged particle bunch or ponderomotive force of the laser field. This forms a ion cavity - commonly referred to as a bubble - that is devoid of electrons; expelled electrons are then attracted back towards the axis forming a plasma sheath around the bubble (see~\cref{fig:1} for a comprehensive sketch). \corr{The described regime is intrinsically non-linear, and lies beyond the reach of complete analytical theories. To accurately predict the blowout structure, and all of the observables of interest to wakefield acceleration, one must therefore resort to large scale kinetic simulations~\cite{esarey-2009,lindstrom-2025}. In this context, Lu \& co-workers~\cite{lu-2006,lu-2006-b} proposed a \textit{phenomenological} model that has become a cornerstone of the plasma-acceleration community, thanks to its strong predictive power and its ability to yield reliable scaling laws. The model, given some input parameters pertaining to the width of the electron sheath, provides an exact second-order differential equation for the bubble radius $r_b(\xi)$ with $\xi=z-\beta_b c t$ the co-moving coordinate ($\beta_b c$ is the speed of the propagating bunch/laser and $z$ the coordinate along its propagation axis; again, see~\cref{fig:1} for a comprehensive sketch). Solving this equation delivers reliable predictions of the shape of the ion cavity and the associated accelerating fields, without resorting to computationally demanding simulations.
Important works in the field rely on this model to support their theoretical analyses. For instance, in high-quality acceleration experiments~\cite{litos-2014}, it is used to compute the optimal \textit{beam loading} for the trailing bunch that enables a flattened accelerating gradient and thus uniform energy gains for both electron~\cite{tzoufras-2008,tzoufras2009beam} and positron acceleration~\cite{zhou-2025}; 
when intense laser pulses drive the wake, a self-injected electron beam may form during the plasma evolution~\cite{malka-2002,pukhov-2002,tsung-2006} In this context, the model predicts the matching conditions required for stable laser propagation by determining the bubble shape~\cite{lu-2007}; in high repetition rate applications, where ion motion influences the plasma recovery process~\cite{darcy-2022}, the model has also been employed to support experimental data.}\\ 
In the last 20 years, the model has been extended and improved in different ways: including the interaction of the electron bunch with the nonlinear plasma wave~\cite{tzoufras2009beam}, considering different forms of the electron sheath~\cite{yi2013analytic,golovanov2016generalised}, including transverse plasma inhomogeneities~\cite{golovanov2016beam,thomas2016non}, considering multi-sheath plasma electron profiles~\cite{dalichaouch-2021}, including bubble excitation~\cite{golovanov2021excitation} and also reformulating the theory based on the energy conservation law~\cite{golovanov-2023}. All these studies, however, have been performed in the cold limit, i.e., by neglecting thermal energy of plasma electrons. This choice is well motivated by the fact that the initial electron thermal energy in the plasma is expected to be of the order of $\corr{k_B} T_{i} \sim 10\; \rm{eV}$~\cite{anania-2014,gonsalves-2019}, which is orders of magnitude smaller than the electron rest energy $m_e c^2 = 0.511$~MeV, resulting in an initial dimensionless temperature $\mu_i = \corr{k_B} T_i / (m_e c^2) \approx 10^{-5}$.
Nevertheless, a complete theoretical framework cannot omit these contributions.
Historically, thermal effects have been studied as a possible regularization mechanism for singularity of the wakefield near the wavebreaking point~\cite{katsouleas-1988, rosenzweig-1988, schroeder-2005, jain-2015}; importantly, recent research has significantly boosted renewed interest in plasma acceleration processes incorporating non-negligible thermal effects: heat accumulation may become a critical factor to control in high-repetition-rate operation~\cite{darcy-2022}; in strongly nonlinear wakes, it has been shown that ion heating on the order of tens to hundreds of keV can counteract on-axis ion pinching, thereby enhancing wake stability~\cite{gholizadeh-2011}; then, following wave breaking, the energy released into the plasma was shown to drive electrons heating up to a few keV, influencing ion motion and enabling soliton formation~\cite{zgadzaj-2020,khudiakov-2022}. For positron acceleration, near-hollow channels with warm electrons (from sub-$\rm{keV}$ to ${\cal O}(10)\;\rm{keV}$) help suppress de-focusing forces and sustain stable propagation~\cite{silva-2021,diederichs-2023,diederichs-2023-b,cao-2024}; lastly, light based diagnostics due to heating may offer new strategies to probe wake evolution~\cite{lee-2024}.
This also prompted the development of novel numerical models to account for thermal effects ~\cite{parise-2022,simeoni-2024,simeoni-2024-b,simeoni-2025}, and  motivates this current study that aims to revisit the model originally proposed by Lu \& coworkers~\cite{lu-2006,lu-2006-b} to include thermal effects. We derive the relevant equations for the bubble radius $r_b(\xi)$ in the presence of a non-negligible thermal background, which converge to the original equations provided by Lu in the limit $\mu_i \rightarrow 0$. These equations are then solved with a suitable boundary condition to account for non-zero thermal spread in the plasma electrons and the results are compared with particle in cell (PIC) numerical simulations. \\ 
The paper is organized as follows. In~\cref{sec:warm-lu-model}, the theory for the warm electron bubble is presented. Comparisons with numerical simulations are given in~\cref{sec:setup}. Conclusions will follow in~\cref{sec:conclusions}.
%
\section{
Theory for warm electron bubble
\label{sec:warm-lu-model}
} 
%
In this section, we show all the calculations that lead to the non-linear ordinary differential equation for the bubble radius $r_b(\xi)$. The derivation shown here closely follows previous calculations~\cite{lu-2006,lu-2006-b}, departing from it only for the introduction of thermal effects. \corr{We remark that the physical framework in which the model is formulated does not rely on a fluid description, nor a mesoscopic one, but relies instead on tracking single representative plasma electrons, treating the dynamics from a Lagrangian, single-particle perspective. The derivation proceeds as follows: first, the equation of motion for a representative electron describing the radius of the bubble is examined under quasi-static, cylindrical, axially symmetric, and complete blowout assumptions. Then, the peculiar geometry of the bubble in the blowout regime is used to give a phenomenological estimate of the electromagnetic sources inside Maxwell's equations, which are solved to obtain the electromagnetic fields, to be then plugged as a Lorentz force into the single electron equation of motion. The resulting equation is the $\textit{Blowout Radius Equation}$, an ODE for $r_b(\xi)$ determining the bubble shape. In the cold limit, this recovers the Lu equation; in the warm case, additional thermal effects act on the initial conditions of the electron and modify force balance.~\cite{jain-2015,diederichs-2023}}. For the rest of this work, we will assume dimensionless variables using as reference quantities the electron's mass $m_e$, its absolute charge $e$ and the speed of light $c$. Consequently, relativistic momenta and energies will be normalized w.r.t. $m_e c$ and $m_ec^2$. Furthermore, number densities will always be given divided by the initial uniform rest number density of plasma electrons $n_0$, equal (due to charge neutrality) to the constant plasma ion's number density $n_i$: ions are considered immobile in our treatment. \corr{Consequently, current densities will be given divided by $cen_0$.} 
In this way, time is made dimensionless through $\omega_p^{-1}$, with $\omega_p =\sqrt{\frac{n_0 e^2}{m_e \epsilon_0}}$ the cold plasma frequency, and lengths are made dimensionless through $k_p^{-1}$, with $k_p=\omega_p/c$ the cold plasma wave-number. Furthermore, for the sake of simplicity, we will focus here on a PWFA application, where the plasma response is induced by a bi-Gaussian beam of electrons moving at the speed of light in the $-\hat{z}$ direction (from now on we assume $\beta_b=-1$) and with longitudinal/radial widths $\sigma_z/\sigma_r$. In co-moving coordinates, its normalized number density $n_b$ reads as:
\begin{align}\label{eq:gaussian-bunch}
    n_b(\xi,r) &= \alpha \exp\left(-\frac{(\xi-\xi_0)^2}{2\sigma_z^2}\right)
    \exp\left(-\frac{r^2}{2\sigma_r^2}\right) \; ,
\end{align}
with $\alpha$ the normalized peak density amplitude, and $\xi_0$ its longitudinal co-moving origin. The whole procedure can easily be adapted to LWFA cases~\cite{lu-2006} where a laser source replaces this electron beam and the perturbing mechanism is due to ponderomotive force, since all the information pertaining the nature of the plasma perturbation is encoded in one single term of the $r_b(\xi)$ equation, as it is pointed out at the end of~\cref{subsec:electromagnetic-fields}. \corr{Throughout this work, we will be expressing temperature using alternatively its dimensionless form (when convenient in formulas) and its physical form (mainly in figures, to give relevant points of comparison to the reader).}

\subsection{
Single electron equation of motion
\label{subsec:single-eq-motion}
}
%
\begin{figure}[h!]
    \centering
    \includegraphics[width=\columnwidth]{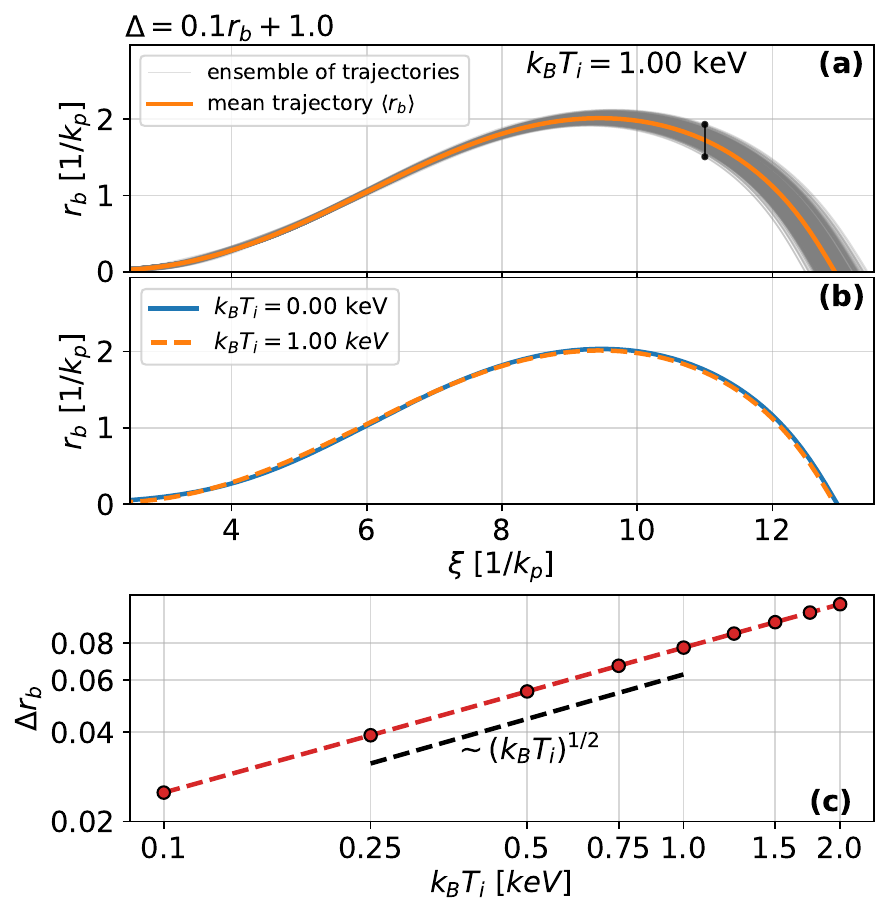}
    \caption{
    \corr{Numerical solutions of~\cref{eq:blowout-radius-equation} obtained using $\Delta=0.1r_b+1.0$.
    Panel a): representative warm plasma solution for $k_B T_i = 1\; \rm{keV}$, showing $500$ stochastic realizations of the boundary-electron trajectories (gray) and their mean $\langle r_b \rangle$ (orange).
    Panel b): comparison between the mean warm trajectory (orange, dashed) and the cold-plasma solution.
    Panel c): measure of the fluctuations $\Delta r_b(\xi)$, evaluated at $\xi=11.0$ (black segment in panel a), plotted versus different background temperatures on a logarithmic scale.}
    }
    \label{fig:2}
\end{figure}
The starting point in the derivation is given by the Maxwell and charge continuity equations for electromagnetic potentials $\phi$ and $\mybf{A}$, normalized w.r.t. $m_ec^2/e$ and $m_ec/e$, and that we express here making use of the Lorenz gauge~\cite{jackson-1998}:
\begin{align}\label{eq:me-and-charge-continuity}
    \begin{cases}
        \partial_t \phi + \nabla \cdot \mybf{A} &= 0 \\
        \partial_t \rho + \nabla \cdot \mybf{J} &= 0
    \end{cases} \; , \quad
    \begin{cases}
    \left( \partial_{t}^2 - \nabla^2 \right) \phi     &= \rho      \\
    \left( \partial_{t}^2 - \nabla^2 \right) \mybf{A} &= \mybf{J}
    \end{cases} \;,
\end{align}
where $\corr{\rho=1-n_e-n_b}$ and $\mybf{J} = - n_e \mybf{\beta} + n_b \hat{\mybf{z}}$ are the electromagnetic field sources, with $n_e$ the plasma electron number density and $\mybf{\beta}$ its velocity. As is customary, one then adopts the co-moving variables $(x,y,z,t)\rightarrow(x,y,\xi=z-ct,t)$ and assumes the validity of the quasi-static approximation~\cite{sprangle-1990,whittum-1997} (QSA, $\partial_t \ll \partial_{\xi}$). Under this assumptions,~\cref{eq:me-and-charge-continuity} becomes:
\begin{align}\label{eq:maxwell-qsa}
    \begin{cases}
    \nabla_{\bot} \cdot \mybf{A}_{\bot} &= \partial_\xi \psi \\
    \nabla_{\bot} \cdot \mybf{J}_{\bot} &= \partial_\xi  S
    \end{cases} \; , \quad
    \begin{cases}
    \nabla^2_{\bot} \psi            &= - S                \\
    \nabla^2_{\bot} \mybf{A}        &= - \mybf{J}
    \end{cases} \; ,
\end{align}
where we have defined the \textit{pseudo-potential} $\psi = -(\phi + A_z)$ and the \textit{source term} $S = -(\rho + J_z)$, and where the differential operator $\nabla_{\bot}$  acts only on the transverse components of the fields (it does not include longitudinal derivatives).\\
The focus is now moved to the motion of a single plasma electron under the action of the electromagnetic potentials previously highlighted. Evolution equations for the relativistic normalized momentum $\mybf{p}=\gamma \mybf{\beta}$ and energy $E = \gamma$ \corr{(with $\gamma=(1-\mybf{\beta}^2)^{-1/2}$ the Lorentz factor)} of said electron are given by~\cite{mora-1997}:
\begin{align}\label{eq:single-electron}
    \frac{d (\mybf{p}-\mybf{A})}{dt} 
    &= \nabla \phi - \mybf{\beta} \cdot (\nabla \mybf{A})
    \; , \\
    \frac{d E}{dt} 
    &= \mybf{\beta} \cdot (\nabla \phi + \partial_t \mybf{A}) \; .
\end{align}
We remark that, at the time-scales of interest, the plasma can be regarded as collisionless. Accordingly, the only force acting on the electron in~\cref{eq:single-electron} is the Lorentz force. Pressure contributions associated with a finite thermal spread in the electron momenta do not appear at the single-particle level: they emerge only as collective effects when the dynamics is described at the hydrodynamic scale~\cite{simeoni-2024-b,schroeder-2010,shadwick-2004}.
The imposition of the QSA, together with the choice of an ultra-relativistic bunch, allows to extract from~\cref{eq:single-electron} a conservation equation in the form~\cite{mora-1997}:
\begin{align}\label{eq:qsa-invariant-conservation}
    \frac{d (p_z+E+\psi)}{dt} = 0 \; .
\end{align}
Hence, we choose an initial point along this trajectory (denoted by the superscript $^{(0)}$ for all related quantities in subsequent formulas) located ahead of the bunch described in~\cref{eq:gaussian-bunch}.
In this unperturbed state, the plasma is globally neutral and at rest, so that no electromagnetic fields are present (i.e., $\mybf{A}^{(0)}=\mybf{0}$ and $\phi^{(0)}=0 \rightarrow \psi^{(0)}=0$). This remains true also in the presence of a finite temperature: although electrons exhibit thermal velocity spreads, their mean momentum is still zero, and the macroscopic fields depend only on average charge and current densities, which still vanish.
One then has:
\begin{align}\label{eq:qsa-invariant}
    p_z + E + \psi = constant = p_z^{(0)} + E^{(0)} \; .
\end{align}
\cref{eq:qsa-invariant} was also derived in~\cite{jain-2015, diederichs-2023} when considering warm plasmas, under considerations similar to the ones presented here. Electron's momentum $p_z^{(0)}$ and energy $E^{(0)}$ can be statistically sampled from an equilibrium Maxwell-Boltzmann
distribution\footnote
{
    In the relativistic regime, the correct momentum distribution is the Maxwell-J{\"u}ttner distribution~\cite{juettner-1911}. However, for a gas at rest and sufficiently low temperatures ($\mu_i \ll 1$), this distribution reduces to the classical Maxwell-Boltzmann form that we are using in this work.
}
with variance $\corr{k_B} T_i$.
By combining~\cref{eq:qsa-invariant} with the mass-shell condition $E^2-p_z^2-\mybf{p}_\perp^2=1$, one can obtain a formula for both the energy and the relativistic longitudinal momentum of the electron:
\begin{equation}\label{eq:momentum-energy-single-particle}
\begin{aligned}
    p_z &= \frac{1 + \mybf{p}_\perp^2 - (\psi-E^{(0)}-p_z^{(0)})^2}{2(\psi-E^{(0)}-    p_z^{(0)})} \; ,  \\
    E   &= - \frac{1 + \mybf{p}_\perp^2 + (\psi-E^{(0)}-p_z^{(0)})^2}{2(\psi-E^{(0)}-p_z^{(0)})} \; ,
\end{aligned}
\end{equation}
and obtain its longitudinal velocity through $\beta_z=p_z/E$. Note that~\cref{eq:momentum-energy-single-particle} represents one of the main differences w.r.t. the cold treatment given in~\cite{lu-2006,lu-2006-b}, and one can recover the original formulation by assuming no thermal background at the initial position of the electron: $p_z^{(0)}=0$ and $E^{(0)}=1$ in dimensionless variables.

From now on, axial symmetry is assumed: we force no dependency on the azimuthal coordinate when moving to a cylindrical $(r,\varphi,z)$ vector basis, and no angular motion for the single electron (i.e., $\beta_\varphi=0$). This last assumption is based on the fact that due to axial symmetry this azimuthal velocity is conserved and expected to be small).
With this assumption, the transversal component of the momentum equation in~\cref{eq:single-electron} assumes then the form~\cite{jain-2015}:
\begin{align}\label{eq:single-electron-radial}
   \frac{dp_r}{dt} &= \partial_r \phi + \partial_t A_r +\beta_z (\partial_z A_r - \partial_r A_z) \; .
\end{align}
\cref{eq:single-electron-radial} can be used to write the leading formula that will translate into a differential equation for the bubble trajectory $r_b(\xi)$. 
Under QSA, and considering the co-moving position of the electron $\xi=\xi[t,z(t)]$, one has:
\begin{align}\label{eq:single-electron-radial-qsa-1}
   \frac{dp_r}{d\xi} &= \partial_\xi A_r - \partial_r A_z - \frac{\partial_r \psi}{1+\beta_z} \; .
\end{align}
Noting that:
\begin{align}\label{eq:pr-transformation}
    p_r = \gamma \beta_r = \gamma \frac{dr}{dt} 
                         = \gamma \frac{dr}{d\xi} \frac{d\xi}{dt}
                         = \gamma \frac{dr}{d\xi} (1+\beta_z) \; ,
\end{align}
one can make use of~\cref{eq:momentum-energy-single-particle} to find:
\begin{align}
    p_r                 &=
    \frac{dr}{d\xi} (E^{(0)}+p_z^{(0)}-\psi) \; , \\
    \frac{1}{1+\beta_z} &=
    \frac{1}{2} \left[ 1+\left(\frac{dr}{d\xi}\right)^2 + \frac{1}{(E^{(0)}+p_z^{(0)}-\psi)^2} \right] \; ,
\end{align}
hence,~\cref{eq:single-electron-radial-qsa-1} becomes:
\begin{equation}\label{eq:single-electron-radial-qsa-2}
\begin{split}
       & \frac{d}{d\xi} \left[ \frac{dr}{d\xi} (E^{(0)}+p_z^{(0)}-\psi) \right] = 
   \partial_\xi A_r - \partial_r A_z \\
   -  &\frac{\partial_r \psi}{2} \left[ 1+\left(\frac{dr}{d\xi}\right)^2 + \frac{1}{(E^{(0)}+p_z^{(0)}-\psi)^2} \right]\; .
\end{split}
\end{equation}

\subsection{Electromagnetic fields \& sources
\label{subsec:electromagnetic-fields}}
%
\begin{figure}
    \centering
    \includegraphics[width=\columnwidth]{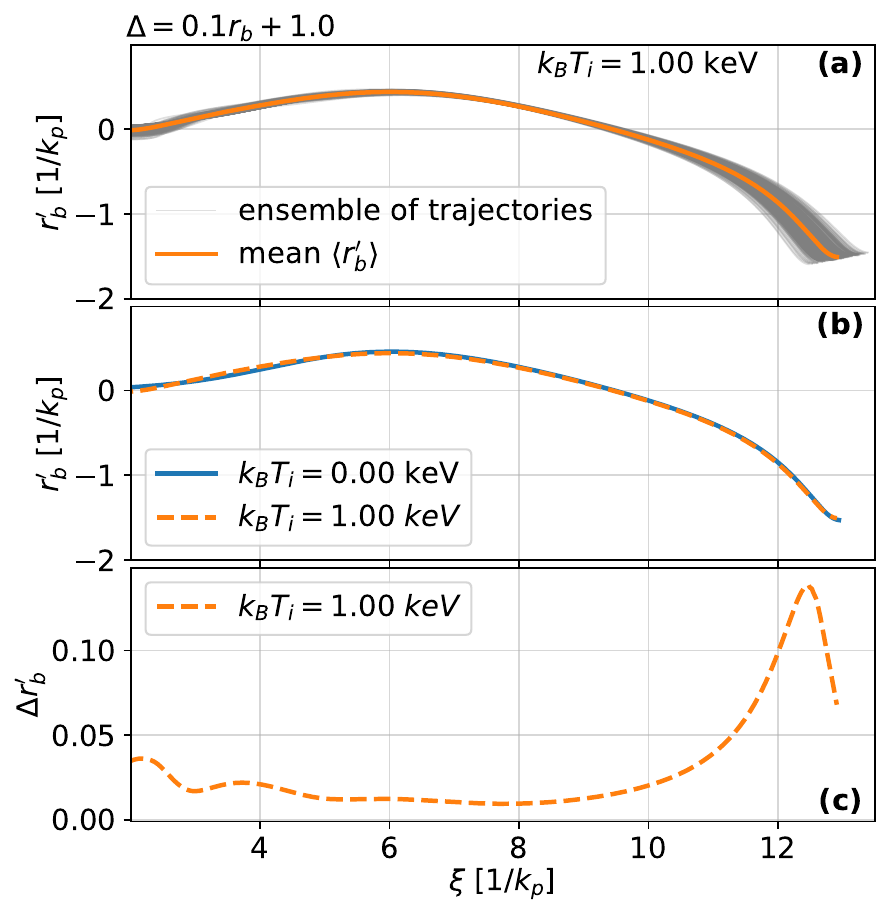}
    \caption{\corr{Analysis of the first derivative $r_b'$ for an initial plasma temperature of $k_B T_i = 1.00 \; \rm{keV}$.
    Panel a): ensemble of $500$ derivative trajectories (gray) and mean $\langle r_b' \rangle$ (orange).
    Panel b): comparison between the mean warm trajectory (orange, dashed) and the cold-plasma solution.
    Panel c): fluctuations $\Delta r_b'(\xi)$, defined analogously to~\cref{eq:thermal-fluctuations}}}
    \label{fig:3}
\end{figure}
\begin{figure}
    \centering
    \includegraphics[width=\columnwidth]{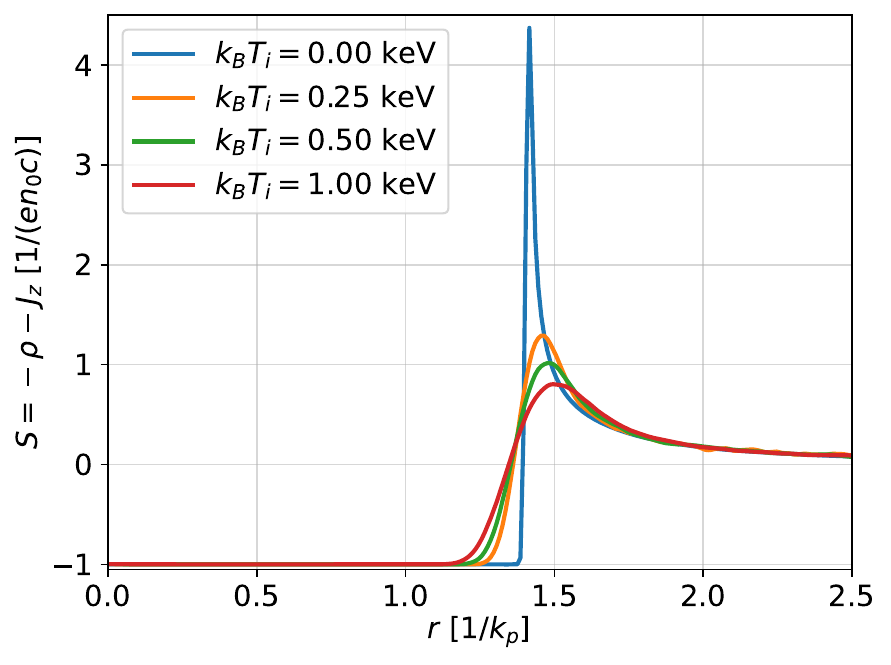}
    \caption{Radial cross section of the source term $S=-\rho-J_z$ for different temperature realizations, at a fixed value of the co-moving coordinate $\xi_0 = 7.0$, corresponding to the longitudinal center of the Gaussian bunch.}
    \label{fig:4}
\end{figure}

\cref{eq:single-electron-radial-qsa-2} constitutes the backbone of the equation we are seeking; to express the equation in a closed form, we need to identify the contribution due to the electromagnetic fields $A_r,A_z,\psi$. In order to do so, one has to \corr{solve} explicitly~\cref{eq:maxwell-qsa} under axial symmetry\corr{, after plugging in the corresponding electromagnetic sources $S$, $J_r$ and $J_z$. These} cannot be determined by first principles and therefore must be approximated via a model. Here, we will be using the so-called \textit{single-sheath} model developed in~\cite{lu-2006,lu-2006-b}, which uses a radial stepwise approximation of $S$ along the blowout bubble. 
In the hypothesis of a complete blowout, where all the plasma electrons are expelled by the presence of a driving bunch, the authors in~\cite{lu-2006,lu-2006-b} identify three main areas of interest:
\begin{align}
    \text{\textit{ Inside the bubble} - (in)}        \;
    & \left[ \; r \leq r_b \; \right]                      
    \;, \notag \\
    \text{\textit{Electron sheath region} - (mid)}    \;
    & \left[ \; r_b < r \leq r_b + \Delta \; \right]  
    \;, \notag \\
    \text{\textit{Unperturbed region} - (out)}        \;
    & \left[ \; r > r_b + \Delta \; \right]                 
    \; .
\end{align}
Here, $r_b(\xi)$ is the blowout radius at longitudinal position $\xi$, while $\Delta(\xi)$ denotes the thickness of the electron sheath surrounding the bubble.
Note that when the plasma temperature is non-zero ($\corr{k_B} T_i \neq 0$) the distinction among the three zones becomes less pronounced  (see, for example,~\cref{fig:1,fig:4}), and a more refined parameterization (like, for example~\cite{dalichaouch-2021}) should be adopted. Nonetheless, for this preliminary study, we maintain the division into these three zones. Subsequent adjustments will be necessary to refine this approach for thermal scenarios.

\corr{In a full blowout scenario, all plasma electrons have been expelled from within the bubble, leaving only bunch electrons, hence the electromagnetic sources are perfectly known therein; One can then obtain the electromagnetic potentials from Maxwell's equations:}
\begin{align}\label{eq:potentials-in}
   \psi^{(in)}              &= \psi_0 + \frac{1}{4} r^2         \; , \notag \\
    A_r^{(in)}              &= \frac{r}{2} \partial_\xi \psi_0  \; ,        \\
    \partial_r A_z^{(in)}   &= - \frac{\lambda}{r}              \; , \notag 
\end{align}
where $\lambda = \int_0^{r} n_b(\xi,r') dr'$. Note that this term will represent the forcing contribution from the trailing bunch in the $r_b(\xi)$ differential equation; therefore, inserting a laser source instead of the one detailed in~\cref{eq:gaussian-bunch} would reduce into properly changing $\lambda$.
Furthermore, when integrating the $r_b(\xi)$ differential equation one will have to impose initial conditions that necessarily start outside of the bubble, even though~\cref{eq:potentials-in} details fields inside the bubble. This inconsistency can be mitigated by assuming a narrow Gaussian bunch (i.e., $\sigma_r \ll 1$), so that the radial position of starting integration point can still be considered ``inside the bunch''. This changes effectively the upper integration point in $\lambda$ to $r \gg \sigma_r$.

\subsection{Blowout radius equation
\label{subsec:blowout-radius-equation}
}
%
We now have all the ingredients for the determination of an equation of motion defining the interface of the bubble. 
If~\cref{eq:single-electron-radial-qsa-2} is evaluated at $r = r_b$, the electromagnetic field expressions from~\cref{eq:potentials-in} can be used, and the only remaining task is to find an appropriate parametrization of $\psi_0$.
This can be done by integrating \corr{Maxwell's equations} after recurring again to the \textit{single-sheath} parametrization of the source term used in~\cite{lu-2006,lu-2006-b}:
\begin{equation}
\begin{cases}
S^{(in)} = -1, & r \leq r_b, \\
S^{(mid)} = n_\Delta, & r_b < r \leq r_b + \Delta, \\
S^{(out)} = 0, & r > r_b + \Delta.
\end{cases}
\end{equation}
where $n_\Delta$ can be obtained from the charge continuity equation. One obtains:
\begin{align}
    \psi_0 &= -\frac{1}{4} r_b^2 (1+\beta) \;, \\
    \beta  &= \frac{2\left(1+\frac{\Delta}{r_b}\right)^2}{\frac{\Delta}{r_b}\left(2+\frac{\Delta}{r_b}\right)}\log\left( 1+ \frac{\Delta}{r_b}\right) - 1 \;.
\end{align}
By assuming a weak dependence of $\Delta$ on $\xi$, meaning that $\partial_{\xi} \beta \sim r_b' \partial_{r_b}\beta$, the blowout radius equation is obtained: 
\begin{align}\label{eq:blowout-radius-equation}
    r_b'' A(r_b) +
    r_b'^{\;2} r_b B(r_b) + 
    r_b C(r_b) = \frac{\lambda(\xi)}{r_b}  \;, 
\end{align}
with
\begin{align}
    \label{eq:bre-coefficients-A}
    A(r_b) &= \left[ E^{(0)}+p^{(0)}_z + r_b^2 \left( 
    \frac{1}{4} + 
    \frac{\beta}{2} + 
    \frac{1}{8} r_b \frac{d\beta}{dr_b} 
    \right) \right] \;, \\
    \label{eq:bre-coefficients-B}
    B(r_b) &= \left[ 
    \frac{1}{2} + 
    \frac{3}{4} \beta + 
    \frac{3}{4} r_b \frac{d\beta}{dr_b} + 
    \frac{1}{8} r_b^2 \frac{d^2\beta}{dr_b^2}
    \right] \;, \\
    \label{eq:bre-coefficients-C}
    C(r_b) &= \frac{1}{4} \left[ 1 + \frac{1}{(E^{(0)}+p^{(0)}_z+\frac{\beta}{4} r_b^2)^2} \right]  \;.
\end{align}
In Lu's model~\cite{lu-2006,lu-2006-b} (that can be recovered by setting $E^{(0)}=1$, $p_z^{(0)}=0$ in~\cref{eq:bre-coefficients-A,eq:bre-coefficients-C}), the evolution of the bubble radius $r_b(\xi)$ is governed by an ordinary differential equation (ODE), whose source term encodes the influence of the electron bunch. In the cold plasma regime, this equation is fully deterministic. However, when thermal effects are introduced, the model becomes stochastic: the ODE's coefficients~\cref{eq:bre-coefficients-A,eq:bre-coefficients-C} become effectively random due to thermal perturbations in the plasma, and this in turn originates into random initial conditions for the ODE. 
As a result, the integration of the ODE must be interpreted over an ensemble of particles, and one has to shift from the deterministic view of the cold model, where the bubble radius $r_b(\xi)$ was representing the "trajectory of the innermost electron", to a warm interpretation where the bubble radius is now a statistical average $\langle r_b(\xi) \rangle$ of an ensemble of randomly sampled trajectories. This stochasticity is driven by the nature of thermal plasma electrons, that even at rest possess non-zero momenta that can be sampled from thermal equilibrium distributions. 

\section{
Model Characterization and Comparison with Numerical Simulations \label{sec:setup}
} 
\begin{figure}
    \centering
    \includegraphics[width=\columnwidth]{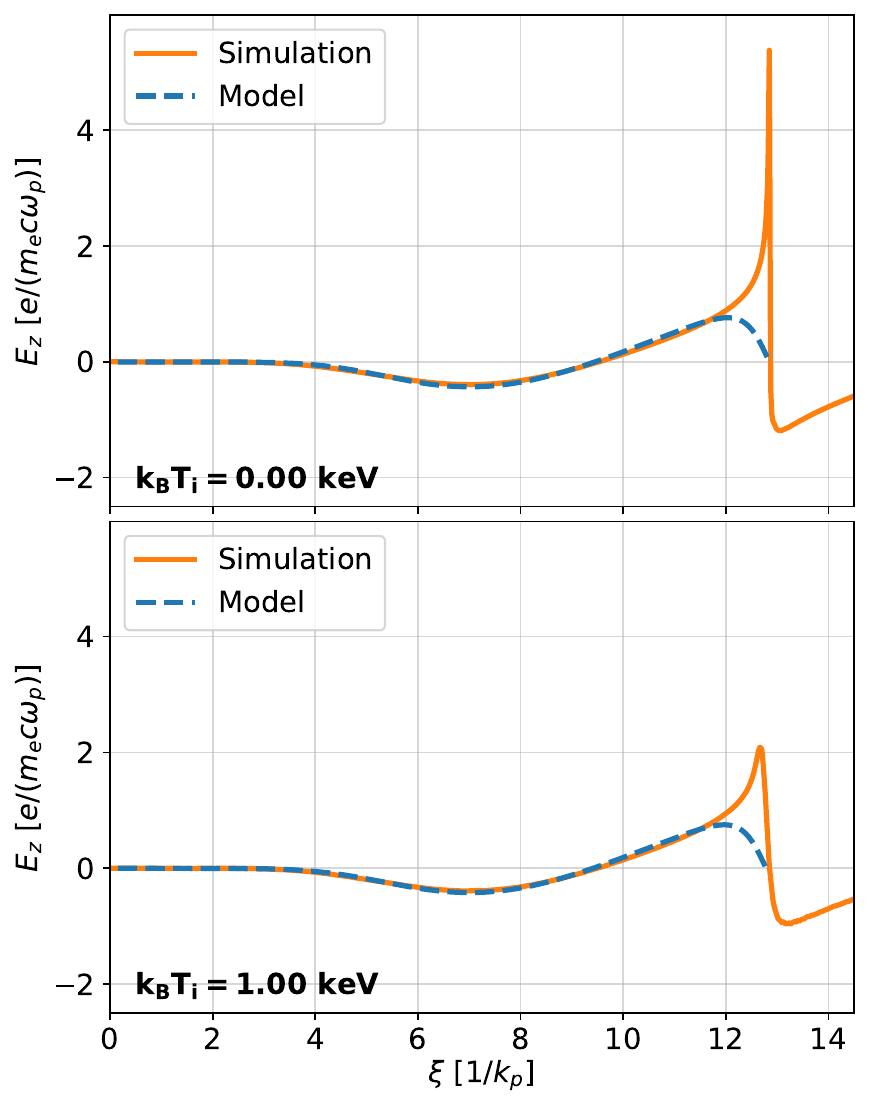}
    \caption{Longitudinal electric fields from PIC simulations (\corr{orange}) compared with model predictions (dashed \corr{blue}) for a cold background plasma (top panel) and for a warm plasma with $\corr{k_B} T_i = 1\; \rm{keV}$ (bottom panel).}
    \label{fig:8}
\end{figure}

\corr{In this section, we will characterize the model by solving the blowout radius equation~\cref{eq:blowout-radius-equation}.}
Furthermore, we will present comparisons with PIC simulations performed using FBPIC~\cite{lehe-2016}, a quasi-3D solver that employs a pseudo-spectral decomposition of the fields in Fourier-Hankel space, enabling tunable azimuthal resolutions, and which also includes standard PIC components  for particle evolution like particle pushers~\cite{ripperda-2018} and charge deposition schemes~\cite{birdsall-2018,esirkepov-2001,steiniger-2023}. All numerical results presented consider a domain of size $L_r=3.0$ and $L_z=15.0$, with cell resolutions $\Delta_z=\Delta_r=0.01$, and computational time step $\Delta t = 0.001$. The background plasma is modeled using \corr{$64$} particles per cell\corr{, and it is initially set at uniform density $n_0=10^{16}\;\rm{cm^{-3}}$. In PIC simulations, the background temperature is implemented through a finite variance in the initial particle velocity distribution}.
We adopt the bi-Gaussian bunch described in~\cref{eq:gaussian-bunch}, with $\alpha=84$, $\sigma_r = 0.1$, and $\sigma_z = \sqrt{2}$, in order to satisfy the narrow bunch condition described at the end of~\cref{subsec:electromagnetic-fields}. The choice of parameters grants a complete blowout regime (the normalized charge parameter
\footnote
{
    For the bi-Gaussian bunch detailed in~\cref{eq:gaussian-bunch}, one defines the normalized charge parameter as $\tilde{Q} = \alpha (2\pi)^{3/2} \sigma_r^2 \sigma_z$~\cite{rosenzweig-2010}.
}
, $\tilde{Q}$, which is often used to evaluate linearity regimes in PWFA applications, equals $\sim 18$) and adhere to the parameters chosen in~\cite{lu-2006,lu-2006-b}.
\begin{figure}
    \centering
    \includegraphics[width=\columnwidth]{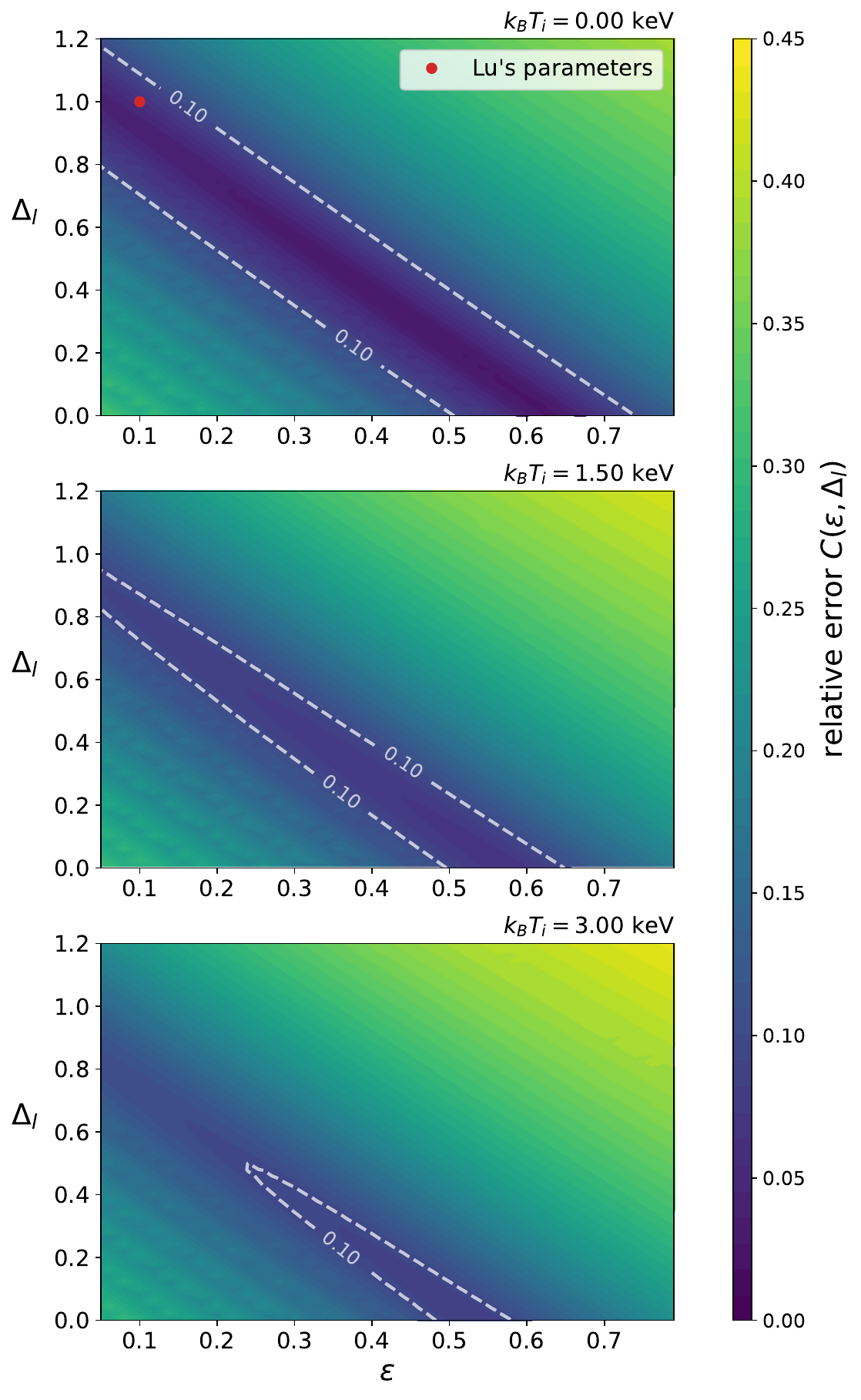}
    \caption{Parametric scan of the electron sheath width parameters $(\epsilon,\Delta_l)$, shown through color maps of the relative error metric $C(\epsilon,\Delta_l)$, as defined in~\cref{eq:cost-function}. Each panel corresponds to a different background temperature: $\corr{k_B} T_i = 0.0,1.5$ and $3.0\; \rm{keV}$ (panels from top to bottom). 
    \corr{Dashed white lines represent $10\%$ error curves.} }
    \label{fig:5}
\end{figure}
In~\cref{fig:1}, we show a snapshot of the system under consideration coming from a PIC simulation. The upper half of the figure corresponds to a cold plasma setup, while the lower half shows the warm plasma case. A clear difference emerges between the two: in the presence of a finite plasma temperature, the plasma bubble becomes longitudinally smaller.
\corr{This reduction in bubble size is expected from linear plasma theory, where finite thermal effects modify the plasma wave frequency through the Bohm-Gross dispersion relation~\cite{bohm-1949,clemmow-1956}, leading to a reduced plasma wavelength~\cite{simeoni-2024,simeoni-2024-b}. In the blowout regime considered here, however, linear arguments do not longer apply, and hence the longitudinal contraction needs to be quantitatively inspected}. 
This effect can be characterized \corr{by looking at} the spatial extent of the interface $r_b(\xi)$, the boundary separating the interior of the bubble from the surrounding plasma sheath, where electrons accumulation occurs. In warm plasmas, this sheath is broader and less sharply defined compared to the cold case~\cite{jain-2015}. This is a direct consequence of the statistical spread in electron momenta due to thermal fluctuations. \\
\begin{figure}
    \centering
    \includegraphics[width=\columnwidth]{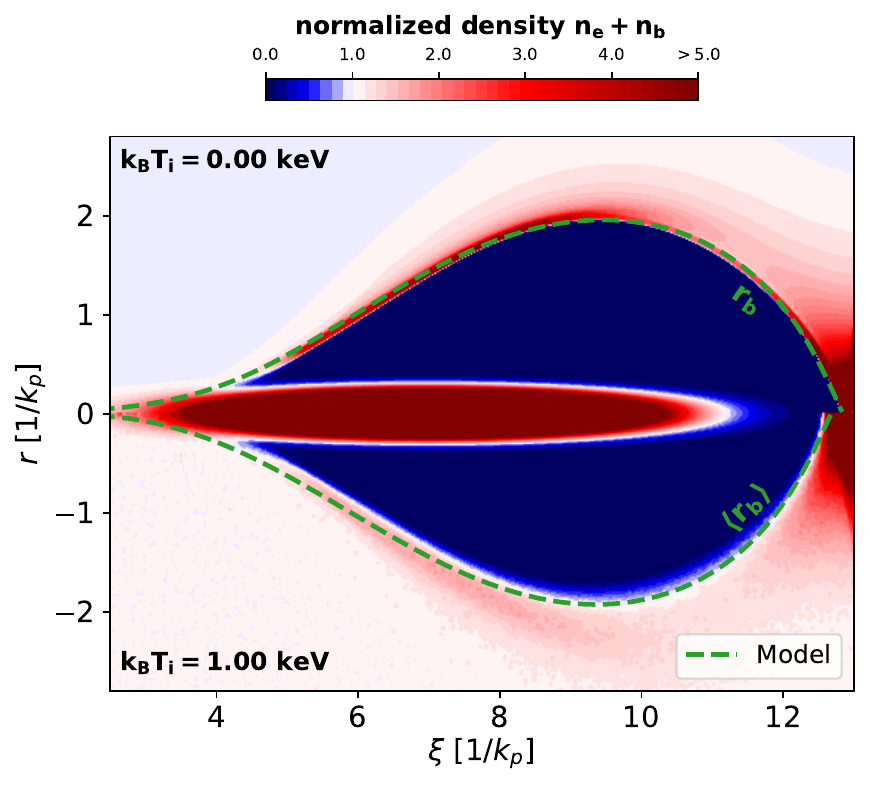}
    \caption{
    \corr{Numerical solutions of~\cref{eq:blowout-radius-equation} compared with PIC results for two initial background temperatures ( $k_BT_i = 0.00\;\rm{keV}$ and $1.00\;\rm{keV}$). In the cold case (top half), the trajectory $r_b(\xi)$ is deterministic; in the warm case (bottom half), we show the ensemble average $\langle r_b(\xi) \rangle$. In both cases, the sheath width $\Delta$ is selected according to the parameter scan of~\cref{fig:5}.}
    }
    \label{fig:6}
\end{figure}
In order to investigate how the theoretical model exposed in~\cref{sec:warm-lu-model} captures this behavior, we numerically solve the ODE in~\cref{eq:blowout-radius-equation} for an ensemble of $500$ stochastic realizations of the initial quantities $E^{(0)}$ and $p_z^{(0)}$. These quantities are sampled according to a thermal distribution, i.e., a normal distribution $\mathcal{N}(\nu,\sigma^2)$ with mean $\nu$ and variance $\sigma^2$, using the relations:
\begin{align}\label{eq:stochastic-sampling}
    p_x^{(0)} &= \mathcal{N}\left(0, \mu_i \right) \;, \notag   \\
    p_y^{(0)} &= \mathcal{N}\left(0, \mu_i \right) \;,          \\
    p_z^{(0)} &= \mathcal{N}\left(0, \mu_i \right) \;, \notag   \\
    E^{(0)}   &= \sqrt{1+\left(p_x^{(0)}\right)^2+\left(p_y^{(0)}\right)^2+\left(p_z^{(0)}\right)^2} \;.                                                \notag
\end{align}
This sampling reflects first on the form of the coefficients of the ODE~\cref{eq:bre-coefficients-A,eq:bre-coefficients-C}. Furthermore, each sampled state yields a different initial condition for the trajectory solver. From~\cref{eq:pr-transformation} one gets:
\begin{align}\label{eq:stochastic-initial-velocity}
    \frac{dr_b}{d\xi}^{(0)} = \frac{p_r^{(0)}}{E^{(0)}\left(1+\beta_z^{(0)}\right)} \;.
\end{align}
Then, for each initial non-zero temperature we compute a {\it statistical average} of the bundle $\langle r_b(\xi) \rangle$.
The effect of this averaging procedure on the theoretical model is depicted in~\cref{fig:2}. Provisionally, we have adopted a fixed electron sheath thickness of the form 
\begin{align}\label{eq:delta-params}
    \Delta  = \Delta_s     + \Delta_l 
            = \epsilon r_b + \Delta_l 
            = 0.1 r_b + 1.0 \;,
\end{align}
using the same parameterization as in Lu's model~\cite{lu-2006,lu-2006-b}, in order to evaluate temperature effects independently of possible changes due to a different choice for the source term $S$. \corr{We remark that in~\cref{eq:delta-params}, $\epsilon$ represents the $r_b$-dependent part of the electron sheath, while $\Delta_l$ is its constant part, according to Lu's own definition.}
In panel (a) of~\cref{fig:2}, we show the solution of~\cref{eq:blowout-radius-equation} for an initial plasma temperature of $\corr{k_B}T_i = 1 \rm{keV}$. The stochasticity of the initial conditions gives rise to an ensemble of trajectories (shown in gray)\corr{, that represents in this model the sheath thermal spreading also observed in~\cite{jain-2015}, and} from which we define the representative bubble trajectory as their statistical average, $\langle r_b(\xi) \rangle$ (shown in orange). 
In panel (b), we compare results coming from a cold realization of the model with the warm $\corr{k_B}T_i = 1 \rm{keV}$ case. One observes that the statistical average $\langle r_b(\xi) \rangle$ (dashed orange line) shows only modest deviations from the cold bubble trajectory (shown in blue), indicating that the averaging process alone is not sufficient to fully capture thermal effects, at least at the temperature values considered here. Nevertheless, the stochastic fluctuations that arise from the ODE are not negligible. In panel (c) we quantify these fluctuations through
\begin{align}\label{eq:thermal-fluctuations}
    \Delta r_b(\xi) = \sqrt{\langle r_b(\xi)^2 \rangle - \langle r_b(\xi) \rangle^2} \; ,
\end{align}
evaluated at $\xi=11.0$, and computed for different values of the initial background temperature. As anticipated from~\cref{eq:stochastic-sampling}, the fluctuations exhibit a $\sim (\corr{k_B} T_i)^{1/2}$ scaling, reminiscent of the behavior found in the expression for the Debye length. We note however that this analogy is only formal, as Debye shielding appears only in the presence of thermal equilibration, and such effect does not appear in this context.\\
A similar analysis is presented in~\cref{fig:3} for the first derivative $r_b'$, which can be related via~\cref{eq:pr-transformation} to momentum variables, and thus its ensemble fluctuations are naturally connected to thermal spread. Panels (a) and (b) lead to the same conclusions as in the previous figures: with the parametrization of~\cref{eq:delta-params}, the stochastic averages deviate only slightly from the cold solution. On the other hand the fluctuations remain significant at both the head and tail of the bubble, and when plotting the metric $\Delta r_b'(\xi)$, defined analogously to~\cref{eq:thermal-fluctuations}, we observe a trend that echoes the temperature profile reported in mono-dimensional studies~\cite{esarey-2007}, with a rise at the bubble front and an even more pronounced peak at its tail.\\
The presence of thermal fluctuations implies a non-trivial impact on the structure of the electron sheath that defines the edge of the bubble and hence, on the effective source term itself. Such effect is clearly visible in~\cref{fig:4}, where we show results from PIC simulations on the radial cross section of the source term $S$ at a fixed value of the co-moving coordinate, $\xi_0 = 7.0$, corresponding to the longitudinal center of the bi-Gaussian bunch. As the background temperature increases, the region of electron accumulation at the boundary of the bubble broadens and its peak amplitude decreases. Indeed, this behavior suggests that an accurate description of the warm plasma regime requires not only the inclusion of stochastic initial conditions in the coefficients $E^{(0)}$ and $p_z^{(0)}$, but also in a temperature dependent parametrization of the source term that captures the underlying statistical spread induced by thermal effects.
A derivation of such parametrization from first principles is indeed challenging and worthy of future investigations. As a preliminary step in this direction, we have performed a parameter-space scan over the terms $(\epsilon,\Delta_l)$ defined in~\cref{eq:delta-params} to identify, \textit{a posteriori}, the most suitable representation of the electron sheath width $\Delta$, and hence of the source term $S$.
To this end, we have introduced a metric that quantifies the discrepancy between the bubble trajectory $r_b^{(sim)}$ obtained from PIC simulations and a theoretical realization $\langle r_b \rangle (\epsilon,\Delta_l)$ of the model~\cref{eq:blowout-radius-equation}, corresponding to a given set of parameters:
\begin{align}\label{eq:cost-function}
    C(\epsilon,\Delta_l) = \frac{||\langle r_b \rangle(\epsilon,\Delta_l) - r_b^{(sim)}||_{\ell_2}}{||r_b^{(sim)}||_{\ell_2}} \;.
\end{align}
The results of this parametric scan are shown in~\cref{fig:5} for three values of background temperature, $\corr{k_B}T_i = 0.0, 1.5, 3.0 \; \rm{keV}$. Comparisons of the $10\%$ error curves (depicted as dashed white lines in the figure) show that an increase in temperature implies a change in the optimal values of both $\Delta_l$ and $\epsilon$.
The \textit{single-sheath} parametrization of the source term adopted by~\cite{lu-2006,lu-2006-b} and in this work implies that the contribution of the source term $S$ to the pseudo-potential $\psi$ can be effectively divided into three distinct regions: an inner cavity, a narrow electron sheath, and an outer linear-response region where plasma perturbation is minimal. One then sees that as $\corr{k_B} T_i$ increases, the $10\%$ error curves marking the optimal parameters region shifts downward. This shows that the optimal $\Delta_l$, expected to be on the order of the cold plasma wavelength $k_p^{-1}$ in the cold limit (equal to $1$ in dimensionless coordinates~\cite{lu-2006,lu-2006-b}), decreases. This trend can be interpreted as a consequence of the overall compression of characteristic length scales in warm plasmas compared to cold ones\corr{~\cite{bohm-1949,clemmow-1956}}.
Additionally, the parameter $\Delta_s=\epsilon r_b$ sets the width of the sheath region, where the source term rises sharply. As shown in~\cref{fig:4}, the height of this region diminishes with increasing temperature; charge conservation then requires a compensating increase in its width. In~\cref{fig:5}, this is reflected into a "thermal shrinkage" of the $10\%$ error curves at low $\epsilon$, effectively excluding these values from the optimal region.
\begin{figure}
    \centering
    \includegraphics[width=\columnwidth]{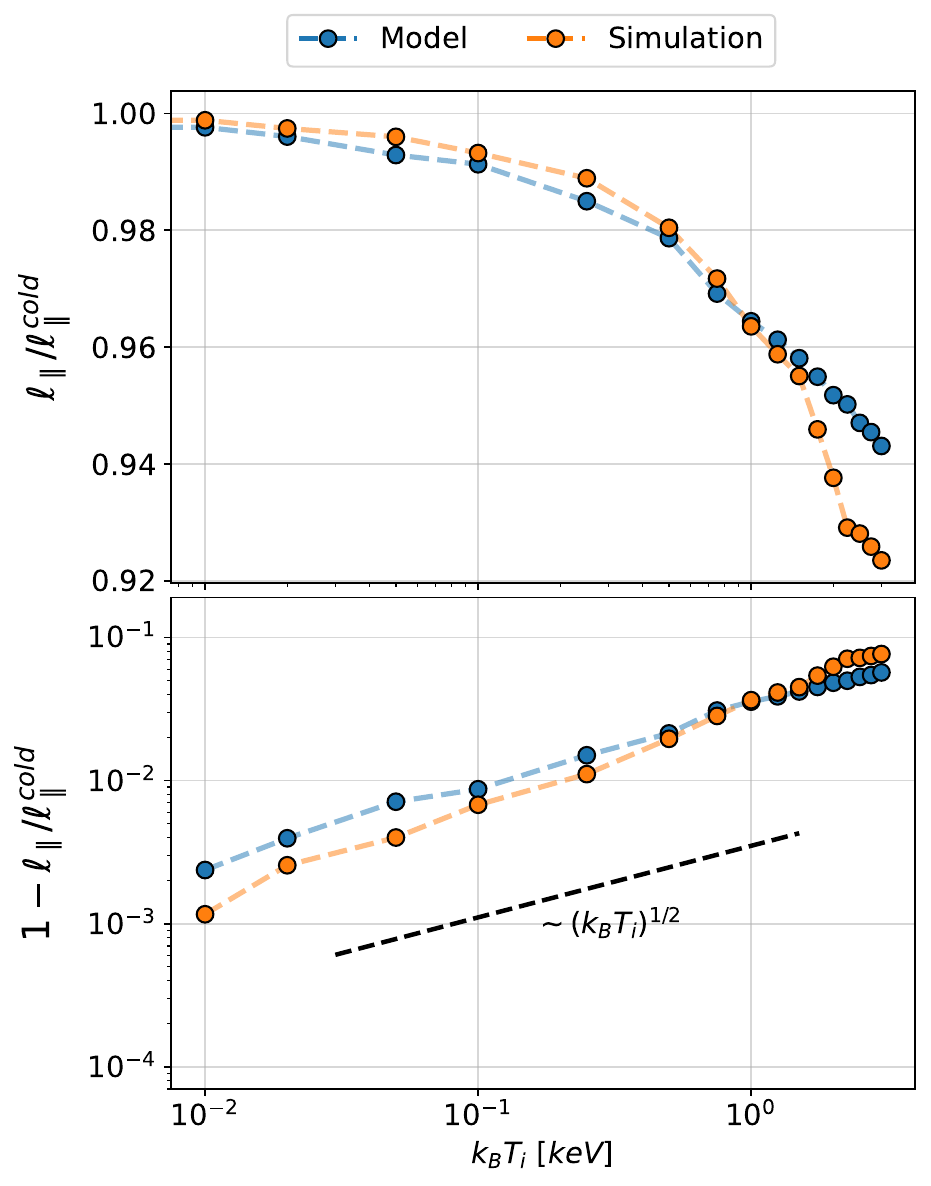}
    \caption{
    \corr{Comparison between PIC simulations (in orange) and the thermal model (in blue). The top panel shows the cold normalized quantity $\ell_\parallel/\ell_\parallel^{cold}$ versus the initial background temperature, while the bottom panel highlights the $\sim (k_B T_i)^{1/2}$ scaling (represented in dashed black as a reference) of the quantity $1-\ell_\parallel/\ell_\parallel^{cold}$}}
    \label{fig:7}
\end{figure}
In \cref{fig:6} we report on the combined effect of the new source term parametrization and the stochastic averaging procedure introduced earlier, showing that they yield a reasonable reconstruction of the interface $r_b$ in the warm-plasma case (bottom half of the figure). In particular, the model successfully captures the reduction in the longitudinal extent of the bubble relative to the cold-plasma case (top half), a consequence of thermal effects.  
\corr{To deepen the analysis, one could quantify how the bubble size varies with temperature to reveal possible scaling laws.  Bubble size can be parametrized via the longitudinal ($\ell_{||}$) and transversal ($\ell_{\perp}$) length scales (see~\cref{fig:1})~\cite{simeoni-2025}. Starting from the maximum radial elongation $r_m$ and the corresponding co-moving coordinate $\xi_m$, the two observables are defined as $\ell_\parallel= \xi_\parallel - \xi_m$ and $\ell_\bot= r_m = r_b\left(\xi=\xi_m \right)$, where $\xi_\parallel$ is the point on the longitudinal axis where the bubble ends. Moreover, since transversal observables are less sensitive to the temperature~\cite{simeoni-2025}, we decided to focus on the characterization of $\ell_{||}$ as a function of $k_{B} T_i$.Results of the analysis are presented in~\cref{fig:7}: in the top panel it is shown that, by carefully selecting the sheath-width parameters within the $10\%$ uncertainty band shown in~\cref{fig:5}, it is possible to reproduce the PIC results for $\ell_\parallel$ using the model, for all the ranges of temperature considered. Furthermore, the bottom panel shows that both the model and the simulations produce a relative contraction following a scaling law of the form
\begin{align}
   1-\ell_\parallel/\ell_\parallel^{cold} \sim (k_B T_i)^{1/2} \;,
\end{align}
which differs from the contraction one would infer from linear plasma theory~\cite{bohm-1949,clemmow-1956}. This square-root dependence is closely related to the thermal broadening of the electron sheath already observed in~\cref{fig:2}.}\\
We note that the magnitude of the thermal effects on bubble size may increase for more linear regimes (corresponding to smaller values of $\tilde{Q}$). However, in such cases, the assumptions underlying the model presented in this work, namely the requirement of a complete blowout, would no longer hold.
We therefore leave the assessment of this thermal model’s performance in a more linear wakefield regime to future studies.\\
Lastly, in~\cref{fig:8} we report the values of the on-axis longitudinal electric fields, for both a cold (top panel) and a warm ($\corr{k_B} T_i = 1\; \rm{keV}$, bottom panel) background plasma. PIC results are confronted against solutions of the theoretical model presented in this article (that can easily be computed from $r_b'$~\cite{lu-2006,lu-2006-b}). It is observed that the model is able to reproduce accurately the field inside the bubble, with noticeable deviations only at its tail (where the interface $r_b(\xi)$ becomes non derivable). Furthermore, the structure of the longitudinal field is not significantly altered by the warm background plasma, as already expected from previous studies~\cite{jain-2015}.
\corr{A more quantitative analysis of the importance of thermal effects on electromagnetic fields can be found in~\cite{simeoni-2025}.}
%
\section{Outlook and Conclusions \label{sec:conclusions}}
We have studied the impact of finite electron temperature on the structure of the first plasma bubble in the blowout regime of plasma wakefield acceleration. Building upon the cold-plasma model introduced by Lu et al. in~\cite{lu-2006,lu-2006-b}, we developed an extended theoretical framework that incorporates stochastic initial conditions and stochastic ODE coefficients arising from thermal momentum distributions. This extension represents a paradigm shift from the deterministic trajectory $r_b(\xi)$ of the cold-plasma case: the resulting ODE of the theory must be solved for an ensemble of particles, and characterized through stochastic averages and fluctuations. 
This must go together with a re-parametrization of the source term adopted in the model: indeed, as shown by PIC simulations, the region where the source term exhibits a sharp rise from the bubble region to the electron sheath is regularized by a non-zero temperature and requires, for each temperature, novel optimal values of the sheath thickness $\Delta$ that one can identify through an \textit{a posteriori} parametric matching procedure. With this correction, the model manages to qualitatively reproduce the temperature induced contraction of both the longitudinal and transverse bubble sizes.
This work is a first step toward the extension of theoretical models to thermal regimes, and can be improved along several directions. First, the seminal model by Lu et al. in~\cite{lu-2006,lu-2006-b} has since been followed by more recent approaches, such as the \textit{multi-shell} model of Dalichaouch et al.~\cite{dalichaouch-2021} and the energy conservation formulation by Golovanov et al.~\cite{golovanov-2023}. One wonders how thermal effects could be implemented in these frameworks, and whether their warm extensions would improve on the current work. Second, the sheath-width parametrization employed here was based on a manual fitting procedure; more refined strategies, including data-driven or machine learning approaches, could be explored to systematically optimize the model and reduce reliance on manual tuning. \corr{Third, in the presence of temperature effects, one could explore whether part of the parametrization of the electron sheath (e.g., its smoothing and the decrease in the peak) can be predicted from "first principles" starting from the cold data or derived from more continuum models~\cite{simeoni-2024,simeoni-2024-b}.} 

\section*{Acknowledgments}
The authors gratefully acknowledge Fabio Bonaccorso for his technical support. We acknowledge useful discussions with S. Romeo and A. Cianchi. This work was supported by the Italian Ministry of University and Research (MUR) under the FARE program (No. R2045J8XAW), project "Smart-HEART". MS gratefully acknowledges the support of the National Center for HPC, Big Data and Quantum Computing, Project CN\_00000013 - CUP E83C22003230001, Mission 4 Component 2 Investment 1.4, funded by the European Union - NextGenerationEU. Financial support from the project DYNAFLO (CUP E85F21004290005) of Tor Vergata University of Rome is acknowledged.

\section*{Author Declarations}

\subsection*{Conflict of Interests}
The authors declare that they have no conflict of interests.

\subsection*{Author Contributions}

\mybf{Daniele Simeoni}: 
Conceptualization (equal); 
Data Curation (lead); 
Formal Analysis (lead); 
Investigation (lead); 
Software (lead); 
Visualization (lead); 
Writing - original draft (lead).
\mybf{Gianmarco Parise}: 
Formal Analysis (supporting);
Investigation (supporting);
Writing - original draft (supporting).
\mybf{Andrea Renato Rossi}: 
Conceptualization (equal);
Formal Analysis (supporting);
Investigation (supporting);
Visualization (supporting); 
Writing - original draft (supporting).
\mybf{Andrea Frazzitta}: 
Conceptualization (equal);
Formal Analysis (supporting);
Investigation (supporting);
Writing - original draft (supporting).
\mybf{Fabio Guglietta}:
Formal Analysis (supporting);
Investigation (supporting);
Writing - original draft (supporting).
\mybf{Mauro Sbragaglia}:
Conceptualization (equal);
Formal Analysis (supporting);
Investigation (supporting);
Writing - original draft (supporting);
Supervision (lead).


\section*{Data Availability Statement}

The data that support the findings of this study are available from the corresponding author upon reasonable request.

\printbibliography
\end{document}